\documentclass[
    ,final            
  ]
  {aipproc}

\layoutstyle{6x9}

\begin{document}

\title{News from a Multi-Wavelength Monitoring Campaign on Mrk 421}

\author{W.~Cui}{
  address={Department of Physics, Purdue University, West Lafayette, Indiana, 
USA}
}

\author{for the VERITAS collaboration}{
  address={see http://veritas.sao.arizona.edu/VERITAS\_members.html for a 
list of the members}
}

\author{M.~Bla\.{z}ejowski}{
  address={Department of Physics, Purdue University, West Lafayette, Indiana, 
USA}
}

\author{M.~Aller}{
  address={Department of Astronomy, University of Michigan, Ann Arbor, 
Michigan, USA}
}

\author{H.~Aller}{
  address={Department of Astronomy, University of Michigan, Ann Arbor, 
Michigan, USA}
}

\author{H.~Ter\"{a}sranta}{
  address={Mets\"ahovi Radio Observatory, Helsinki University of Technology, 
Espoo, Finland}
}

\author{B.~Mochejska}{
  address={Harvard-Smithsonian Center for Astrophysics, Cambridge, 
Massachusetts, USA}
}


\author{P.~Boltwood}{
  address={1655 Main Street, Stittsville, ON K2S 1N6, Canada}
}

\author{A.~Sadun}{
  address={Department of Physics, University of Colorado at Denver, Colorado, 
USA}
}

\author{M.~B\"{o}ttcher}{
  address={Department of Astronomy, Ohio University, Athens, Ohio, USA}
}

\author{A.~Reimer}{
  address={Department of Astronomy, University of Bochum, Bochum, Germany}
}

\begin{abstract}
We conducted a daily monitoring campaign on Mrk~421 in 2003 and 2004 with
the Whipple 10 m telescope and the large-area instruments aboard {\em RXTE},
simultaneously covering TeV and X-ray energies. Supporting observations at
optical and radio wavelengths were also frequently carried out. Mrk~421
was observed over a wide range of fluxes (with a dynamic range of $\sim$30 
both at TeV and X-ray energies). The source was relatively quiet in 
2003 but became unusually active in 2004, with flares reaching peak fluxes 
of $\sim$80 mCrab in X-rays and $>$3 Crab at TeV energies! We will describe 
the multiwavelength campaign and present some preliminary results. We will 
also discuss the implications of the results on the proposed emission 
models for TeV blazars.
\end{abstract}

\maketitle

\section{Introduction}
Over the past decade, one of the most exciting advances in high energy
astrophysics has been the detection of sources at TeV energies with
ground-based $\gamma$-ray facilities (see Weekes 2003 for a recent review). 
Among the sources 
detected, blazars are arguably the most intriguing. The emission from 
a blazar is generally thought to be dominated by radiation from a 
relativistic jet that is directed roughly along the line of sight 
(Urry \& Padovani 1995). The spectral energy distribution (SED) of TeV
blazars invariably shows two characteristic ``humps'' in the $\nu F_{\nu}$ 
representation, with one located at X-ray energies and the other at TeV 
energies (Fossati et al. 1998). There is a general correlation between 
the two SED peaks. TeV blazars are also known to undergo flaring episodes, 
both at X-ray and TeV energies. The flares have been observed over a wide 
range of timescales, from months down to minutes. The observed X-ray flaring 
hierarchy in Mrk 421 seems to imply a scale-invariant physical origin of 
the flares (Cui 2004).

A popular class of models associates the X-ray emission from a TeV blazar 
with synchrotron radiation from highly relativistic electrons in the jet 
and the TeV emission with synchrotron self-Compton (SSC) processes (e.g., 
Marscher \& Gear 1985; Maraschi et al. 1992). The SSC models can, therefore, 
naturally account for the observed X-ray--TeV correlation. Moreover, they 
have also enjoyed some success in re-producing the observed SEDs. However, 
the models face challenges, such as the presence of ``orphan TeV flares'' 
(in 1ES 1959+650, Krawczynski et al. 2004; and in Mrk 421, see Fig.~1 in 
this work). Alternatively, the jet could be energetically 
dominated by the magnetic field, and it is the synchrotron radiation from 
highly relativistic protons that might be responsible for the observed 
TeV gamma rays (Aharonian 2000; M\"ucke et al. 2003). In such a scenario, 
the X-rays are thought to be mainly due to synchrotron radiation from 
co-accelerated electrons in the jet. Although the proton-synchrotron models 
can also describe the observed SEDs and accomodate the X-ray--TeV 
correlation, they are being challenged by variability timescales below
$\sim$15 min in TeV blazars. Very recently, it has been proposed that the 
flares in blazars might be associated with magnetic reconnection events in 
a magnetically dominated jet (Lyutikov 2003), perhaps similar to solar 
flares in this regard. The model might offer a natural explanation for the 
hierarchical flaring phenomenon in Mrk 421, in analogy to solar flares. 
However, it has not been applied to the data in any quantitative manner.

To make further progress on distinguishing emission models proposed for 
TeV blazars, we believe that simultaneous or contemporaneous data are 
critically needed over a wide range of fluxes, especially in the crucial 
X-ray and TeV bands, for quantifying the SED and its variability of a 
source and for allowing investigations of such important issues as 
variability timescales, cross-band correlation, spectral variability and 
hysteresis, and so on. Such data are severely lacking at present, despite 
intense observational efforts over the years.

\section{The Monitoring Campaign}

We conducted a comprehensive multi-wavelength monitoring campaign on 
Mrk 421. It is the first known TeV blazar and remains to be the only
one that can be detected nearly all the time at TeV energies. The
campaign started in the late Febuary of 2003 and lasted until the early 
May of 2004. The effort was strongly focused on the X-ray and TeV bands. 
We took a snapshot of Mrk 421 at X-ray energies with the PCA detector on 
{\em RXTE} (with typical exposure times of 2--3 ks) and at TeV energies
with the Whipple 10 m telescope (with comparable exposure times), which
were coordinated for each night during a Whipple observing period.
Supporting observations were also carried out, but less frequently, at
4.8, 8.0, and 14.5 GHz with the 26 m UMRAO telescope at the University of 
Michigan, at 22 and 37 GHz with the 14~m Mets\"ahovi Radio Telescope at 
the Helsinki University of Technology, and at optical wavelengths with the 
1.2 m telescope (BVRI photometry) at the Fred Lawrence Whipple 
Observatory (FLWO) and the 0.4 m telescope (R band only) at the 
Boltwood Observatory. The Boltwood flux measurements are differential; 
the results have been normalized to those of the FLWO measurements, 
thanks to a significant overlap of the two observing programs. We note 
that the error bars shown for the Boltwood data do not include systematic 
uncertainties and are thus certainly an under-estimation of the overall 
measurement error. No optical data are currently available for the 
Whipple 2002/2003 observing season.

\section{Preliminary Results}

In this section, we present and discuss some of the results from the
observations, which we must stress are still preliminary in nature.

\subsection{Light Curves}

The X-ray and TeV light curves from the entire campaign are shown in 
Fig.~1, along with radio and optical ones in the selected bands. Note 
that most Whipple observations were conducted at small zenith angles 
($< 30^{\circ}$). We found that variations in the zenith angle hardly 
cause any noticeable effects in the TeV light curves shown. In general,
Mrk 421 was relative quiet in 2003, although it was still seen to vary
significantly at X-ray and TeV energies. The measured X-ray and TeV 
count rates are roughly correlated but are clearly not always in step
(note the presence of ``orphan'' TeV flares). The latter is even more 
obvious in 2004, when the source became unusually active. Many 
major flaring episodes were observed then. For instance, a 
giant outburst took place near the end of the campaign, with the source 
reaching peak fluxes of $\sim$80 mCrab in X-rays and $>$3 Crabs in the 
TeV band, respectively. It is worth noting that during the giant outburst 
the TeV emission appears to reach the peak much sooner than the X-ray 
emission. The light curves also show that Mrk 421 varies progressively 
less towards longer wavelengths.
\begin{figure}[hp]
\caption{Preliminary light curves of Mrk 421. For clarity, the light 
curves are shown separately for the Whipple 2002/2003 and 2003/2004 
observing seasons. For the R-band data, the FLWO points are shown in
bullets and the Boltwood points in diamonds. Note that no optical data 
are available for the 2002/2003 season; the 22 GHz data are shown 
instead. }
\includegraphics[height=.3\textheight]{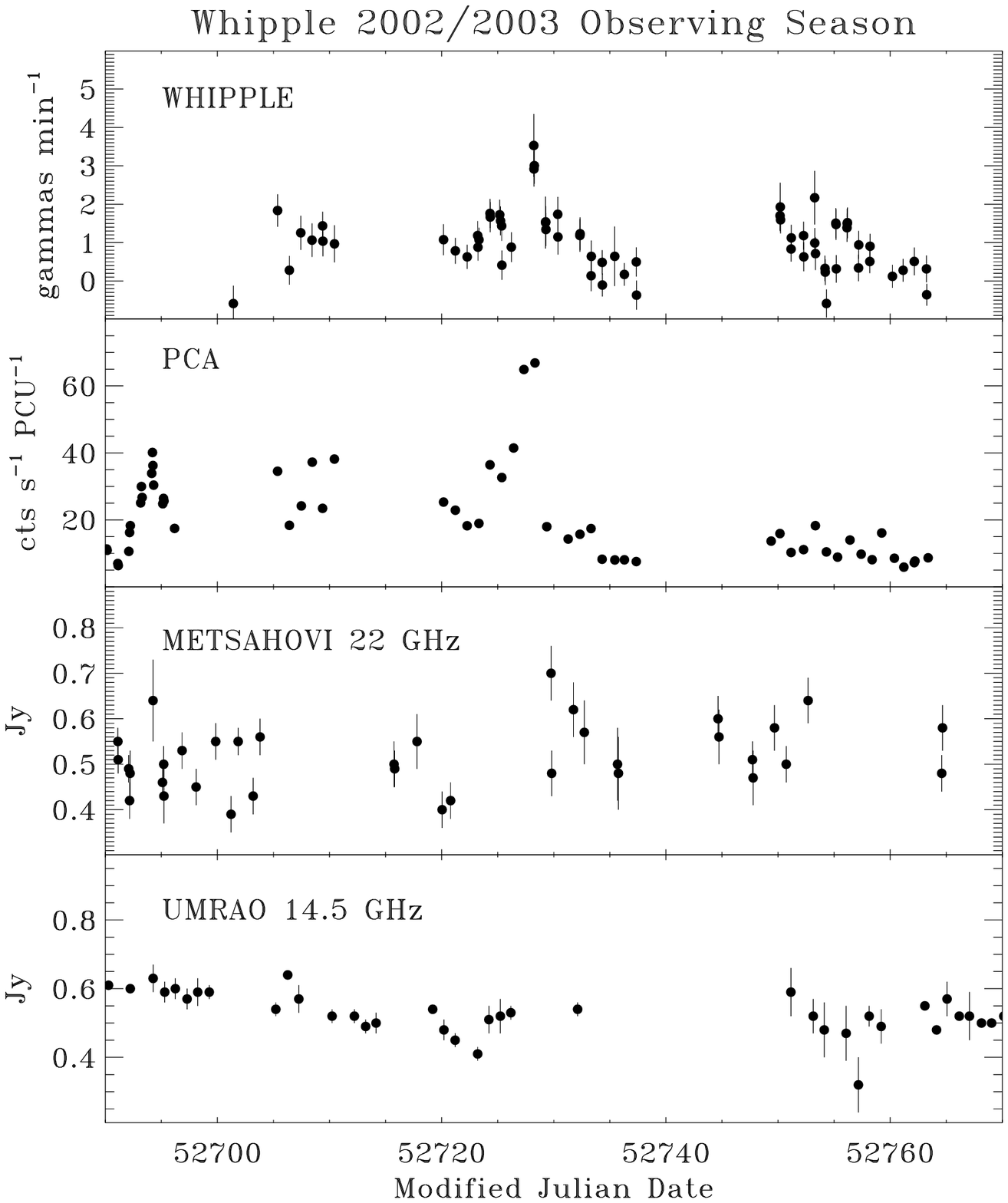}
\includegraphics[height=.3\textheight]{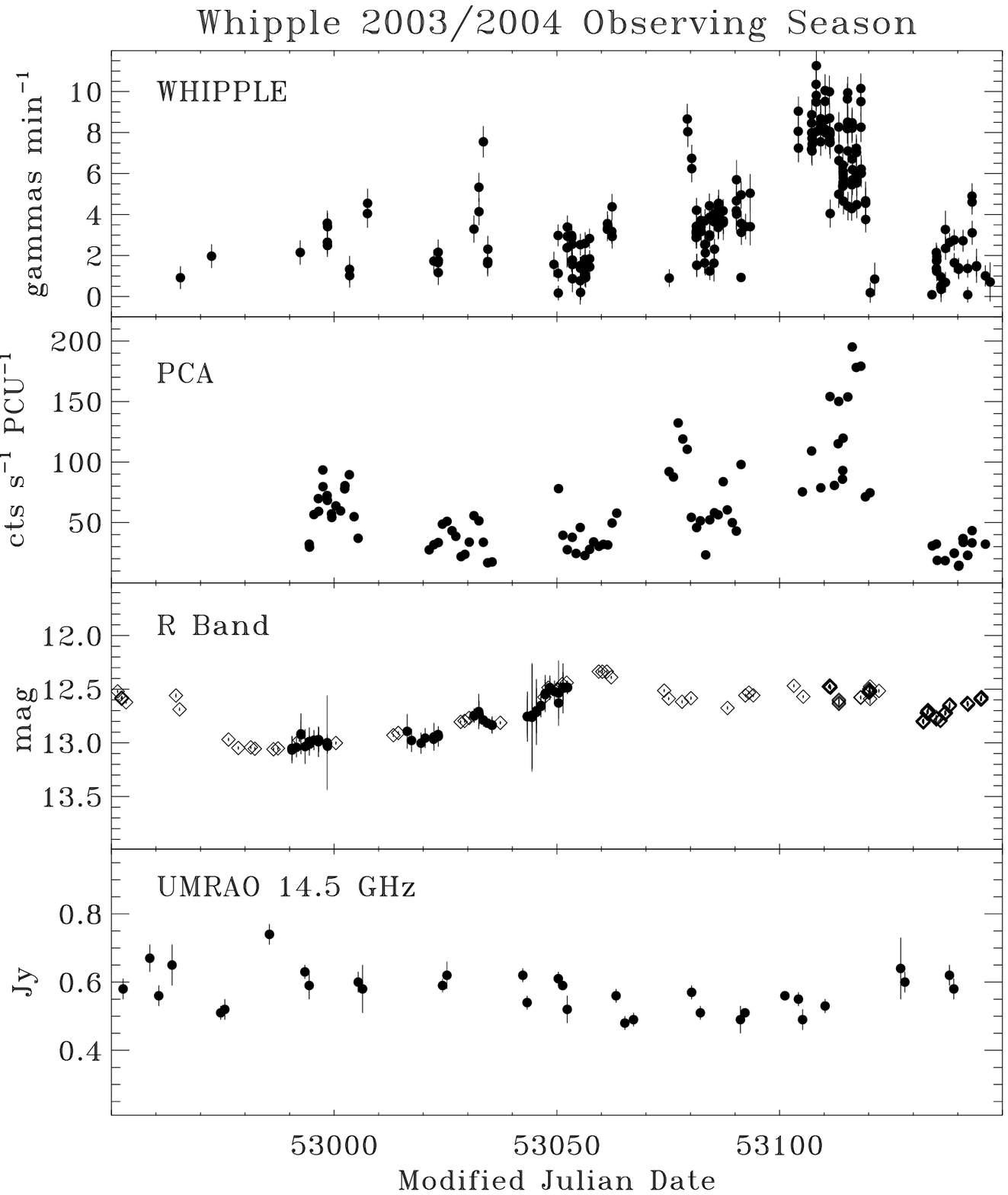}
\end{figure}

\subsubsection{X-ray--TeV Correlation}

The correlation between the X-ray and TeV emission can be examined more
quantitatively by plotting the count rates against each other, as shown
in Fig.~2. We can see that the dynamical range of the data collected is
quite large, about a factor of 30 in both energy bands, which is 
important for studying correlative variability of Mrk 421. The correlation 
is clear from the figure, but it is only a loose one, unlike
what one might naively expect based on the SSC scenarios. 
To be more rigorous, we computed the 
cross-correlation function between the X-ray and TeV light curves, using 
the interpolation method, and the results for the 2003/2004 data are shown 
in Fig.~2. There are two main peaks in the correlation function: one is 
centered roughly at zero lag and the other at +180 hours. We caution, 
however, that correlation functions are, in general, prone to systematic 
effects such as irregular data gaps. An investigation of such effects on
our data is under way. If confirmed, the positive lag would imply that 
on average TeV photons lead X-ray photons, which is opposite to what SSC 
models predict. Looking at the data more carefully, we noticed that the 
peak at the positive lag is dominated by data taken during the giant 
outburst.
\begin{figure}[hp]
\caption{{\em (left)} Relationship between the X-ray and TeV count rates 
of Mrk 421. The solid line shows the best linear fit. 
{\em (right)} Cross-correlation function between the X-ray and TeV count 
rates of Mrk 421 in the Whipple 2003/2004 observing season. }
\includegraphics[height=.20\textheight]{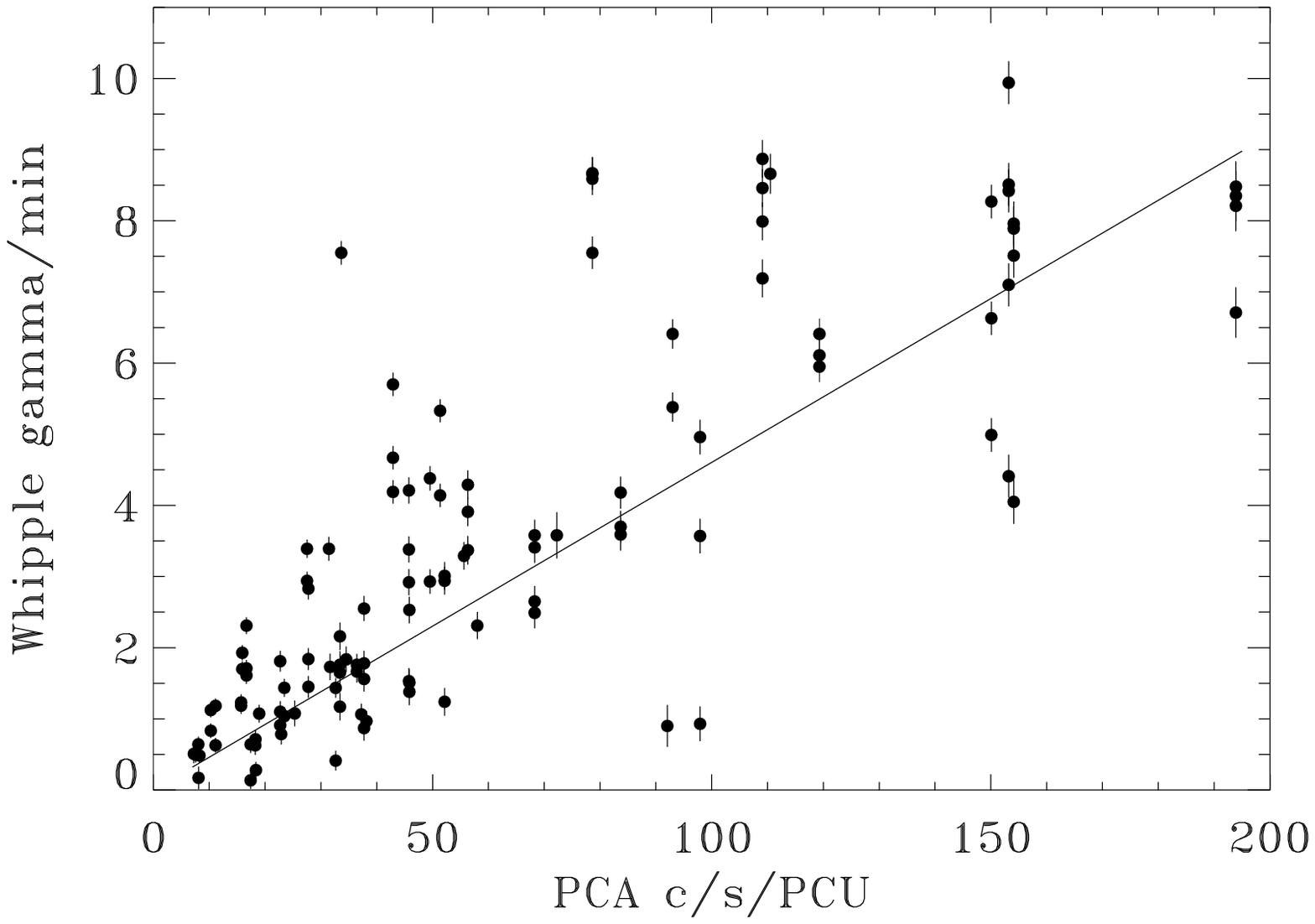}
\includegraphics[height=.20\textheight]{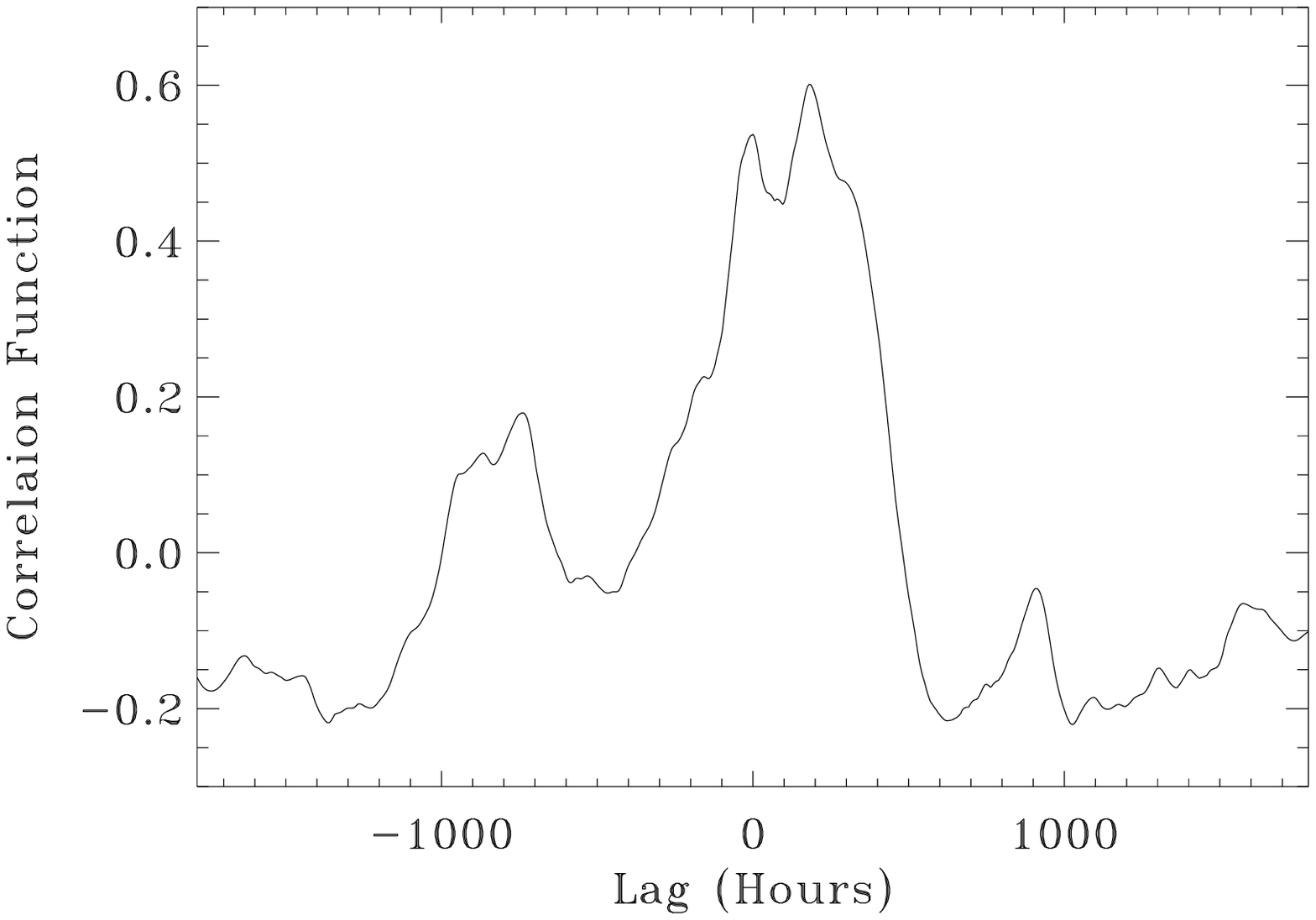}
\end{figure}

\subsubsection{Rapid X-ray and TeV flares}

Zooming onto the X-ray and TeV light curves, we found flares on much
shorter timescales. Fig.~3 shows examples of sub-hour X-ray flares.
The most rapid X-ray flare lasted only for $\sim$15 minutes and shows
substantial sub-structures, implying variability on even shorter 
timescales. Since the X-ray emission is almost certainly of synchrotron 
origin, the timescales associated with such short flares lead to severe 
constraints on the size of the flaring region, the strength of the 
magnetic field, and the Doppler factor of the jet bulk motion in a 
relatively model-independent manner (Cui 2004). 

Only on one occasion, a sub-hour X-ray flare was detected during a TeV 
flare. No counterpart is apparent at TeV energies, as shown in Fig.~3. 
The data do not allow comparisons on longer timescales, which shows a 
serious drawback of the ``snapshot'' 
observing strategy. For studying rapid flaring activities of TeV blazars, 
long, continuous observations are required. This is the 
trade-off that one must consider, however, given the limited resources 
available. We detected a number of rapid TeV flares on timescales of 
hours, all during the period of the giant outburst. Fig.~4 shows a 
sample of such flares. Due to observational constraints (note that the 
coverage was already intensified at TeV energies during the period), 
the flares were only partially covered. Though not statistically 
significant, weaker flares on shorter timescales appear to be present.
\begin{figure}[tp]
\caption{{\em (left)} Most rapid X-ray flare detected in Mrk 421. Note the 
sub-structures in the profile of the flare. {\em (right)} Sub-hour X-ray 
flare during a rapid TeV flare (see Fig.~4). }
\includegraphics[height=.25\textheight]{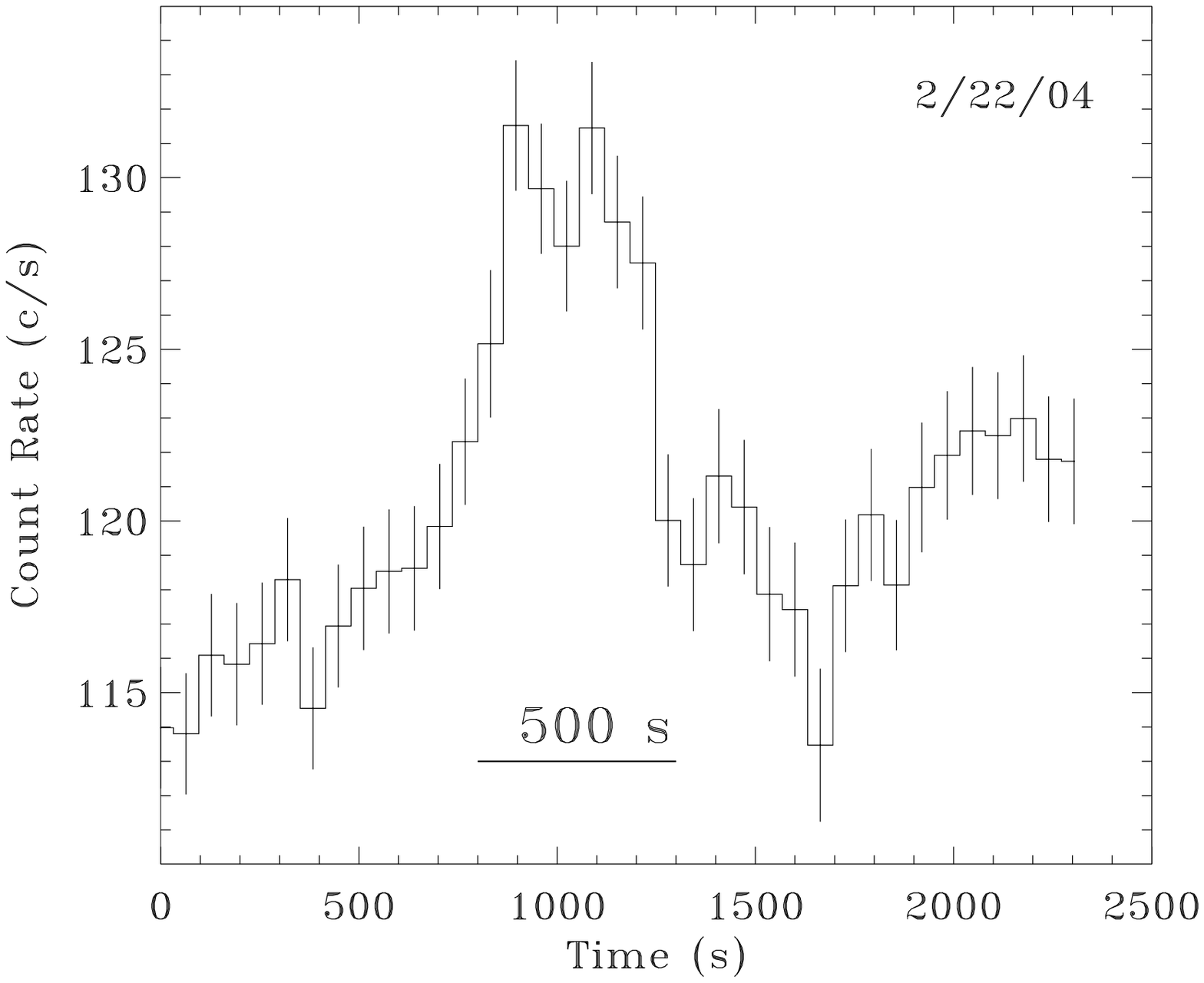}
\includegraphics[height=.25\textheight]{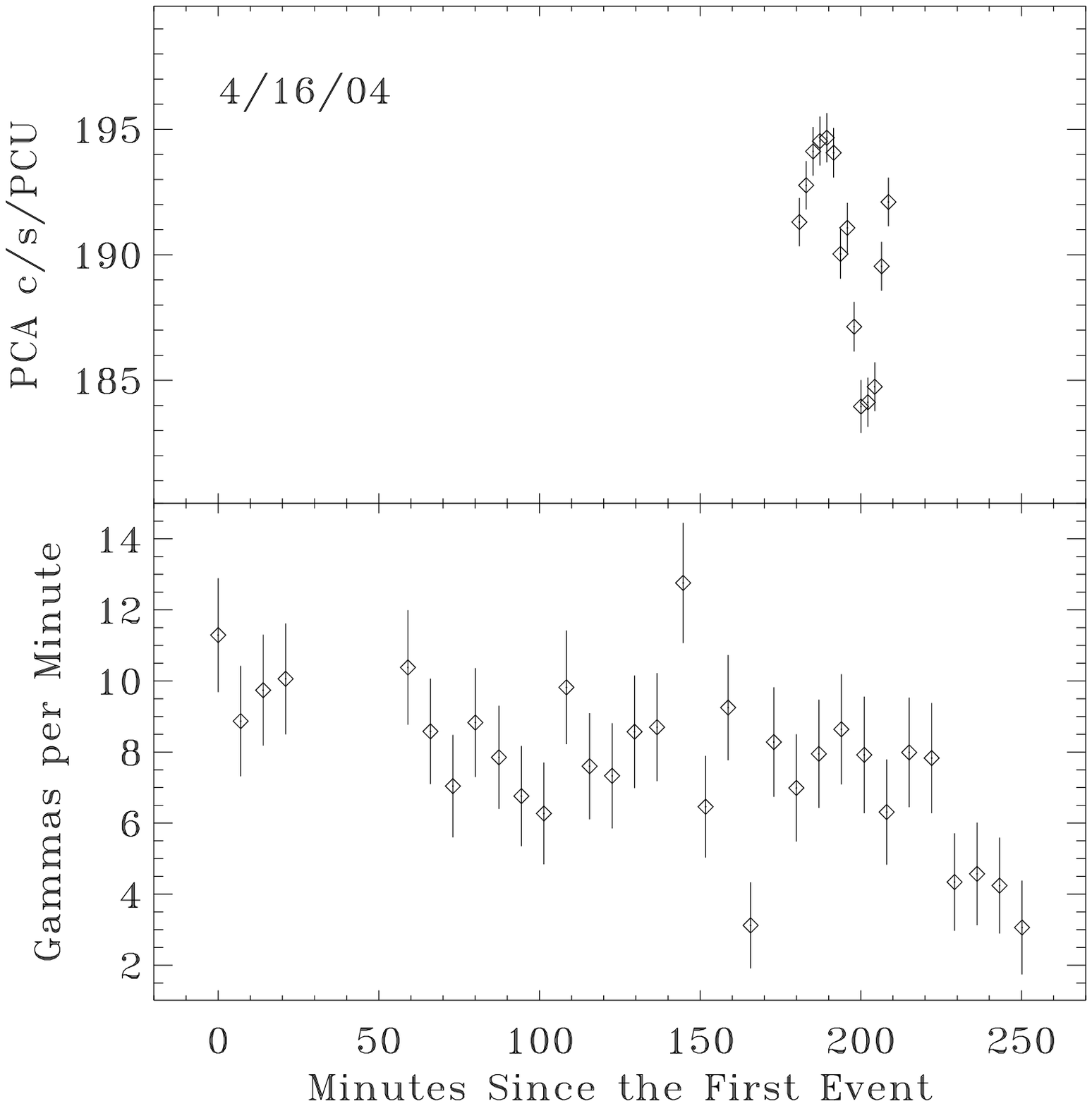}
\end{figure}

\begin{figure}[tp]
\caption{Rapid TeV flares in Mrk 421. }
\includegraphics[height=.49\textheight,angle=90]{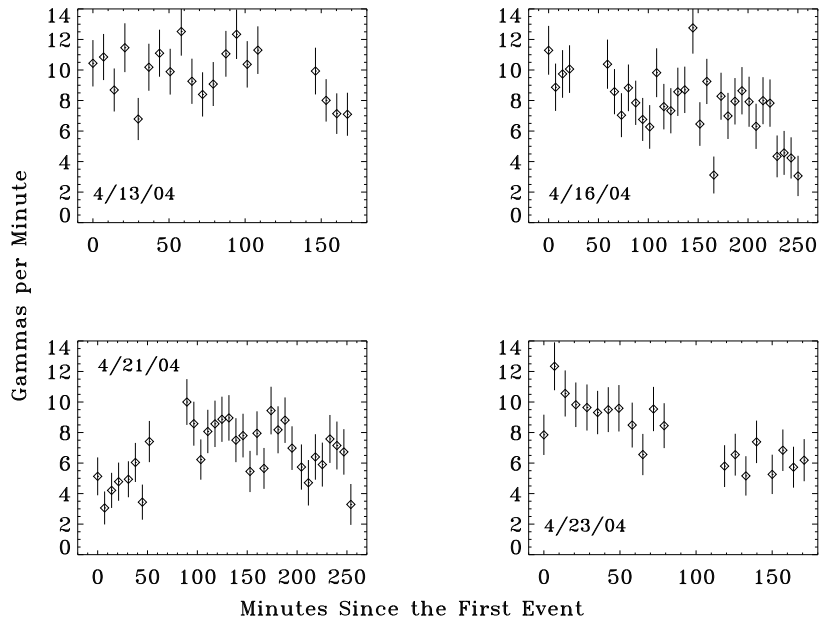}
\end{figure}

\subsection{Spectral Energy Distributions}

To examine spectral variability, we divided the observations into 8 
groups, based on X-ray count rates of Mrk 421. For each group, we 
constructed an SED from radio, optical, X-ray, and TeV observations
in the group. Note that there usually is a much fewer number of 
observations at radio and optical wavelengths, while the observations
at X-ray and TeV energies match reasonably well (by design). Since
Mrk 421 varies little at longer wavelengths, we believe that the
resulted SED should be quite reliable. Fig.~5 shows the SEDs in three
of the groups, with each differing from its neighboring group roughly 
by a factor of 3 in average X-ray count rates.
\begin{figure}[tp]
\caption{Spectral energy distribution of Mrk 421 at different X-ray
fluxes. The X-ray flux differs roughly by a factor of 3 between two
adjacent SEDs. The solid curves are fits to the data with a 
one-zone SSC model (with contributions from the synchrotron 
radiation and SSC shown separately). For comparison, the dot-dashed
line shows the effect of ignoring absorption by the IR background. }
\includegraphics[height=.4\textheight]{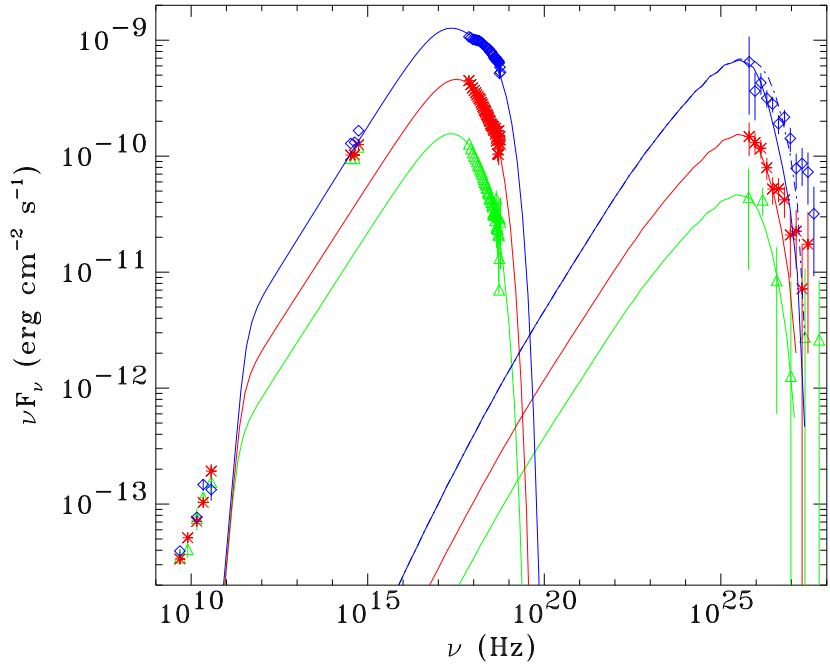}
\end{figure}

The SED varies greatly both at X-ray and TeV energies but little at
radio and optical wavelengths. This is hardly surprising in the 
context of the SSC models, since X-ray and TeV photons originate from 
the same population of most energetic electrons in the jet. The 
synchroton cooling time is much shorter for X-ray emitting electrons 
than for those contributing to radio or optical emission, which might 
account for the difference in variability in different bands. The SED 
is flatter (or harder) at higher fluxes; this is particularly apparent 
at X-ray energies. This is consistent with results from previous works 
(e.g., Fossati et al. 2000; Krennrich et al. 2002; Aharonian et al. 
2002).

We experimented with a one-zone SSC model\footnote{The description of 
the code can be found at http://jelley.wustl.edu/multiwave/spectrum/.} 
to fit the data, taking into account attenuation of TeV fluxes by the 
diffuse infrared background. The results are shown in Fig.~5. The model 
seems to adequately describe the SEDs at X-ray and TeV energies, 
although there are still some discrepancies at TeV energies. On the 
other hand, it fails badly to account for the measured radio and 
optical fluxes. This is a generally known problem with one-zone SSC 
models (e.g., Maraschi et al. 1999; Krawczynski et al. 2004), which is 
hardly surprising since real AGN jets are known to have complicated 
structures. In modeling, the problem is usually dealt with by invoking 
multiple populations of electrons (with different energy distributions) 
in different zones (e.g., Maraschi et al. 1999). The efforts to carry 
out more sophisticated modeling of the data collected in our campaign
are currently in progress.

\section{Concluding Remarks}

A substantial amount multiwavelength data have been collected on 
Mrk 421 through our monitoring campaign. The data have allowed us to
carry out investigations on variability timescales, cross-band
correlation, SED and its flux dependence, and spectral variability
and hysteresis across flares. Some preliminary results have been
presented in this contribution. They have already begun to shed light 
on emission mechanisms in TeV blazars. The study of most rapid flares
is hindered by the snapshot observing strategy adopted. It is,
therefore, highly desirable to take a complementary approach in
the future by observing the source continously for sufficiently 
long time.


\section{Acknowledgments}

The VERITAS collaboration is supported by the US Department of Energy, 
National Science Foundation (NSF), the Smithonian Institution, PPARC (UK), 
NSERC (Canada), and Science Foundation Ireland. W. Cui and 
M. Bla\.{z}ejowski also gratefully acknowledge support from NASA. UMRAO 
is supported in part by funds from the NSF and from the University of 
Michigan Department of Astronomy.

\section{References}
Aharonian, F. 2000, New Astronomy, 5, 377 \\
Aharonian, F., et al. 2002, A\&A, 393, 89 \\
Cui,~W. 2004, ApJ, 605, 662 \\
Fossati, G., et al. 1998, MNRAS, 299, 433 \\
Fossati, G., et al. 2000, ApJ, 541, 166 \\
Krawczynski,~H., et al. 2004, ApJ, 601, 151 \\
Krennrich.~F., et al. 2002, ApJ, 575, L9 \\
Lyutikov,~M. 2003, New Astr. Rev. 47, 513 \\
Maraschi, L., Ghisellini, G., \& Celotti, A., 1992, ApJ, 397, L5 \\
Maraschi, L., et al. 1999, 526, L81 \\
Marscher, A. P., \& Gear, W. K., 1985, ApJ, 298, 11 \\
M\"ucke, A., et al. 2003, APh, 18, 593 \\
Urry,~C.~M., \& Padovani,~P. 1995, PASP, 107, 803 \\ 
Weekes,~T.~C. 2003, Proc. 28th ICRC (astro-ph/0312179) \\

\end{document}